\documentstyle[aps,prl,multicol,psfig]{revtex}

\begin{document}
\draft
\title{Probing superconducting phase fluctuations from the current noise spectrum
of pseudogaped metal-superconductor tunnel junctions}

\author{Xi Dai, Tao Xiang$^{\dagger *}$, Tai-Kai Ng and Zhao-bin Su$^\dagger$}

\address{Physics Department, Hong Kong University of 
Science and Technology, 
Clear Water Bay, Kowloon, Hong Kong}

\address{$^\dagger$Institute of Theoretical Physics, Chinese Academy
of Sciences, P.O. Box 2735, Beijing 100080, the People's Republic of China}

\address{$^*$Isaac Newton Institute for Mathematical Sciences, the University 
of Cambridge, 20 Clarkson Road, Cambridge CB3 0HE, U.K.}

\date{\today}

\maketitle

\begin{abstract}
We study the current noise spectra of a tunnel junction of a metal with
strong pairing phase fluctuation and a superconductor. It is shown that
there is a characteristic peak in the noise spectrum at the intrinsic
Josephson frequency $\omega _J=2eV$ when $\omega _J$ is smaller than the
pairing gap but larger than the pairing scattering rate. In the presence of
an AC voltage, the tunnelling current noise shows a series of characteristic
peaks with increasing DC voltage. Experimental observation of these peaks
will give direct evidence of the pair fluctuation in the normal state of
high-$T_c$ superconductors and from the half width of the peaks the pair
decay rate can be estimated.
\end{abstract}

\input{psfig.sty}

\begin{multicols}{2}

Direct measurements of the pair susceptibility in the normal state of a
superconductor are important for understanding the pseudogap phenomenon in
high-$T_c$ superconductors. For conventional superconductors, the pairing
fluctuation exists only in a very narrow temperature range above $T_c$.
However, in the underdoped high-$T_c$ materials, 
it is believed that the pair fluctuation is
important in a substantial temperature range in the normal state \cite
{Emery95}. This fluctuation is also important in superconductor thin films,
where long ranged phase coherence is strongly suppressed \cite{thin}. The
pseudogap effect was often interpreted as a precursor to the high-$T_c$
superconductivity. It was observed in many different experiments \cite
{nmr1,Mar,Loe,HD1,loram,opt,tun1}. However, what was really found in
experiments is the reduction of the low-lying density of states associated
with the opening of an energy gap, which may or may not be a pairing gap,
and a direct observation of the pair phase fluctuation is still absent.

The pair susceptibility in the normal state of a underdoped cuprate can be
measured from the linear response of the system to an external pair field,
like the measurement of the magnetic susceptibility. Such an external pair
field can be introduced, for example, by coupling this underdoped cuprate to
an optimally doped cuprate with a higher superconducting transition
temperature via a tunnelling junction. When the temperature lies between the
superconducting transition temperatures of these two materials, the
tunnelling process will undoubtedly be strongly affected by the pair
fluctuation in the normal side\cite{pair}.

The Josephson effect occurs when both sides of the junction are
in the superconducting state. A DC voltage across the junction leads to an
oscillating Josephson current at frequency $\omega _J=2eV$, where $V$ is the
applied voltage. If one side of the junction is not in the superconducting
phase, the Josephson current is zero on average. However, the Josephson
effect does not disappear completely. It manifests in the current noise
spectrum and, as will be shown later, the current noise is greatly enhanced
at the Josephson frequency $\omega _J$ if the pairing fluctuation is strong.

In this paper, we present a theoretical analysis of the tunnelling current
noise spectrum of the pseudogaped metal-superconductor junction as 
discussed above. We assume
that the pairing on both sides of the junction has $d_{x^2-y^2}$-wave
symmetry. The extension to other pairing symmetry cases is straightforward.
The main results we obtained are: (1) With an applied DC voltage, there is
a characteristic Josephson peak in the noise spectrum at the intrinsic
Josephson frequency $\omega _J$, although the coherent Josephson current is
zero. (2) When an AC voltage $V_s\cos \left( \omega _st\right) $ is applied
in addition to the DC voltage, there are a series of characteristic peaks in
the zero frequency noise centred at $\omega _J=n\omega _s$ ($n$ an integer)
with increasing $\omega _J$. (3) the half width of the peak is proportional
to the pair decay rate $\Gamma $ of the system, but the peak intensity
drops very quickly with increasing $\Gamma $. 

The Hamiltonian of the junction we study is defined by $H=H_S+H_N+H_T$ with
\begin{eqnarray}
H_S &=&\sum_{p\sigma }\varepsilon _pd_{p\sigma }^{+}d_{p\sigma
}+\sum_p\left( \Delta \varphi _pd_{p\uparrow }d_{-p\downarrow }+h.c.\right) ,
\label{super} \\
H_N &=&\sum_{k\sigma }\varepsilon _kc_{k\sigma }^{+}c_{k\sigma
}+\sum_{kq}\left[ \Psi (q,t)c_{-k+q/2,\uparrow
}c_{k+q/2,\downarrow }+h.c.\right]   \nonumber \\
&&+\sum_q\chi _{{\rm pair}}^{-1}(q,t,t^{\prime })\Psi ^{*}(q,t)\Psi
(-q,t^{\prime }),  \label{normal} \\
H_T &=&\sum_{kp\sigma }V_{kp}e^{i\int_0^teV_s\cos (\omega _st^{\prime
})dt^{\prime }+i\omega _Jt}c_{k\sigma }^{+}d_{p\sigma }+h.c.,  \label{tunnel}
\end{eqnarray}
where $\varphi _p=\cos p_x-\cos p_y$ and $\Psi (x,t)=\Delta _n\exp \left[
i\phi (x,t)\right] $ is the pairing field of the normal lead. $H_S$ is the
Hamiltonian for a $d$-wave superconductor. $H_N$ is the effective
Hamiltonian of a metal with pairing fluctuations \cite{pair,levin1} and $
\chi _{{\rm pair}}$ is the pair propagator (or pair susceptibility) in the
normal state. In $H_T$, $V_{kp}$ is the tunnelling matrix and can be expanded
with the crystal harmonics \cite{pair,hirs}. To the second order
approximation, it is given by
\begin{equation}
|V_{pk}|^2=V_0^2+V_1^2\varphi _k\varphi _p.
\end{equation}
The constant term in $|V_{pk}|^2$ has contribution to the single particle
tunnelling current, but no contribution to the pair tunnelling current because
of the $d$ -wave pairing symmetry. The integration resulted from the applied
AC voltage on the exponent in $H_T$ is difficult to handle analytically. In
practical calculations, it is often expanded as a sum of a serial of
oscillating terms with the Bessel function: 
\[
e^{i\int_0^teV_s\cos (\omega _st^{\prime })dt^{\prime }}=\sum_nJ_n\left( {
\frac{{eV_s}}{{\omega _s}}}\right) e^{in\omega _st},
\]
where $J_n(x)$ is the n'th order Bessel function \cite{mahan}.

Let us first consider phenomenologically 
the effect of pairing phase fluctuation on the current 
noise. We assume
that the phase fluctuation follows a diffusion process and the spatial 
correlation length is very short. 
In this case, the correlation function of the phase field is given
by 
\begin{equation}
\left\langle e^{i\phi (x,t)}e^{-i\phi (x^{\prime },0)}\right\rangle \sim
e^{-\Gamma |t|/2}\delta(x-x'),  \label{pair}
\end{equation}
where $\Gamma $ is the pair fluctuation rate. From this equation, it can be
shown that 
\begin{equation}
\chi _{{\rm pair}}(q,\omega )=1/(i\omega +\Gamma /2).
\end{equation}

The current noise spectrum measures the current fluctuation in the junction.
It is defined by the symmetrized current-current function: 
\begin{equation}
S(\omega )\equiv \int dte^{i\omega t}\left\langle \left\{ I(t)-\left\langle
I(t)\right\rangle ,I(0)-\left\langle I(0)\right\rangle \right\}
\right\rangle ,  \label{sw}
\end{equation}
where $I(t)$ is the current across the junction and $\left\{ A,B\right\}
\equiv AB+BA$. In the case only a DC voltage is applied, the above tunnelling
Hamiltonian can be treated semi-classically and the current-current
correlation function contributed from the pair tunnelling is simply given by
the thermal average of the phase fields:
\begin{eqnarray}
\left\langle I(t)I(0)\right\rangle &\sim &\int dx dx'
\left\langle \sin \left[ \phi
(x,t)+\omega _Jt\right] \sin \phi (x^{\prime },0)\right\rangle
\nonumber \\ 
&\sim & \cos \left( \omega _Jt\right) e^{-\Gamma |t|/2}
\end{eqnarray}
Substituting this equation into (\ref{sw}), we then obtain the noise
spectrum contributed from the pair tunnelling 
\begin{equation}
S_{pair}(\omega )\sim {\frac{{\Gamma }}{{(\omega -\omega _J)^2+\Gamma ^2/4}}}
+{\frac{{\Gamma }}{{(\omega +\omega _J)^2+\Gamma ^2/4}}.}  \label{noise}
\end{equation}
This result indicates that the Josephson current
cross the junction is coherent if the tunnelling time is much shorter than
the fluctuation time $\tau \sim \Gamma ^{-1}$ of virtual cooper pairs. In
the limit $\Gamma \rightarrow 0$, $S(\omega )$ becomes two delta-functions
centred at $\pm {\omega _J}$. For finite $\Gamma $, these two
delta-functions are broadened and the half width of the broadened peak is
proportional to $\Gamma $.

Eq. (\ref{noise}) is similar as the result obtained by Martin and Balatsky
very recently \cite{Ivar}. However, as both sides of the junction in the
experimental setup they suggested are in the normal state, the total pair
decay rate is larger than the case discussed here. Since the
intensity of the peak compared with the average noise drops sharply with
increasing ${\Gamma }$ (Figure 2), it may be 
difficult to observe the
characteristic Josephson fluctuation in finite frequency conductance
measurement suggested in their paper. While in our paper,
to determine the pair decay rate only the 
zero frequency noise is required which is easier to measure. 

In a s-wave pairing system, since no quasiparticle excitations 
exist within the
energy gap, the single particle tunnelling can be ignored when the noise
frequency is lower than the gap. For the d-wave pairing case, the
contribution of the single particle tunnelling cannot be ignored because of
the existence of gap nodes. To find out the contribution from both the
quasiparticle and pair tunnelling to the noise, we 
perform a microscopic calculation the
current-current correlation function using the closed time Green's function
technique\cite{Kel,chou}. The tunnelling Hamiltonian $H_T$ is treated as a
perturbation. Since the normal Josephson effect is absent in the junction,
the second order contribution of $H_T$ contains only the quasiparticle
tunnelling terms. However, the noise caused by the quasiparticle tunnelling
varies very slowly with frequency. When the noise frequency is smaller than $
\Delta $, it adds only a featureless background to the noise spectra.

The most important contribution of $H_T$ to the noise is from the fourth
order perturbation, in particular from the virtual pair tunnelling process.
When the pairing fluctuation in the normal state is strong, a Cooper pair
can tunnel from the superconductor to the normal lead and contributes a
significant term to the noise at low frequencies. Figure 1 shows the Feymann
diagram for this pair tunnelling term in the current-current correlation
function. The solid line is the Green's function of electrons. The dashed
line is the propagator of the pair field in the normal state.

\begin{figure}[htb]
\begin{center}
\psfig{file=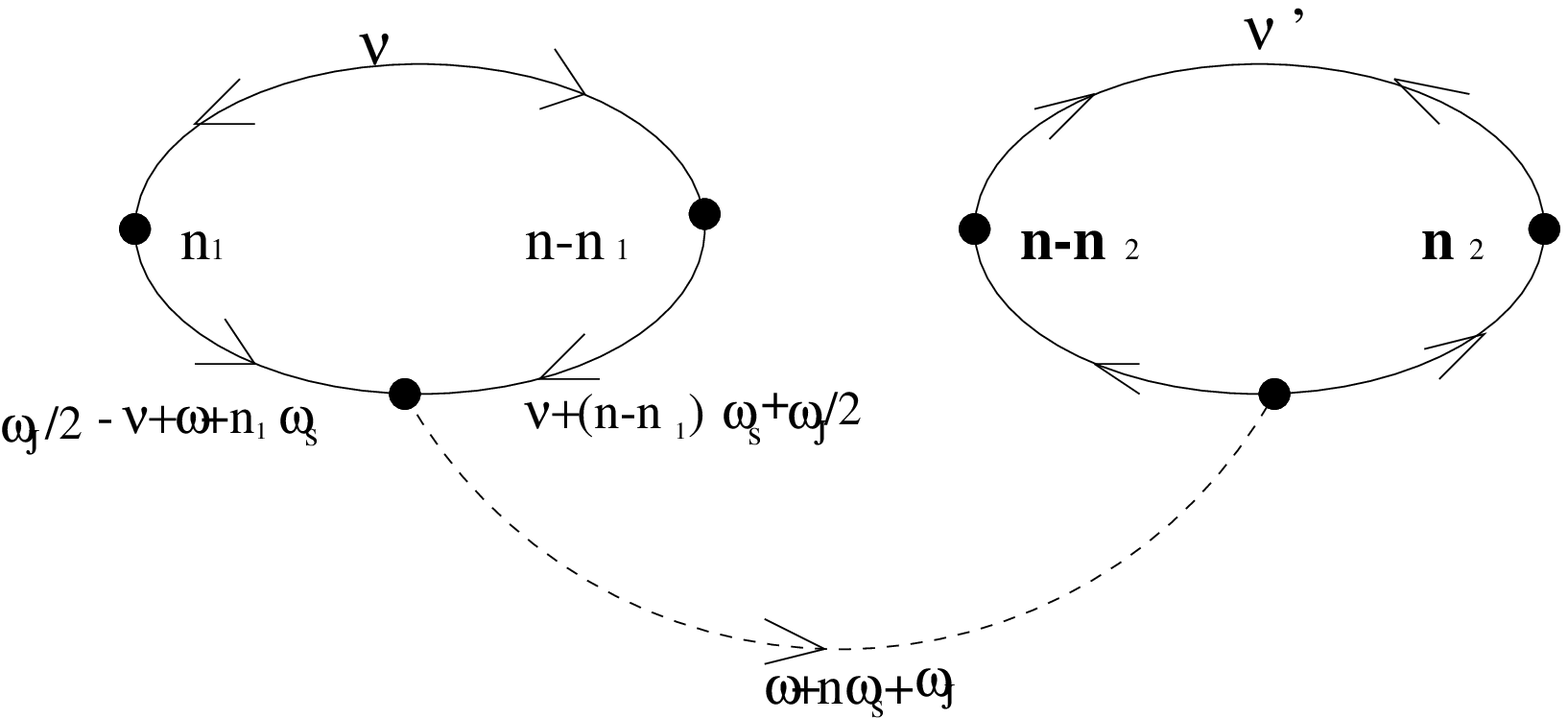,height=40mm,width=85mm,angle=0}
\end{center}
\caption{The Feymann diagram for the pair tunnelling term, where
the solid line represents the electronic Green's function and
the dashed line represents the pairing fluctuation.}
\label{Fig.1}
\end{figure}

To the leading order approximation in both the quasiparticle and pair
tunnelling channels, we find that the noise spectra is given by 
\begin{equation}
S(\omega )\approx N_q(\omega )+N_p(\omega )+N_q(-\omega )+N_p(-\omega ),
\end{equation}
where $N_q$ and $N_p$ are respectively the contributions from the
quasiparticle and pair tunnelling:

\end{multicols}

\begin{center}
--------------------------------------------------------------
\end{center}

\begin{eqnarray}
N_q(\omega ) &=&{V_0^2}\sum_nJ_n^2(x)\rho _s\int d\nu \rho _d(\nu )\left[
n_f(\nu )-n_f(\nu +\omega _n^{\prime \prime })\right] \coth \frac{\beta
\omega _n^{\prime \prime }}2,  \label{nq} \\
N_p(\omega ) &=&V_1^4C\sum_{n,n_1,n_2}\frac{
J_{n_1}(x)J_{n-n_1}(x)J_{n_2}(x)J_{n-n_2}(x)}{{\omega _n^{^{\prime
}2}+\Gamma ^2/4}}\coth \frac{\beta \omega _n^{\prime }}2  \nonumber\\
&&\sum_q\left\{ {\omega _n^{\prime }}\left[ \phi _2^2(\omega _n^{\prime
},q)-\phi _1^2(\omega _n^{\prime },q)\right] +{\Gamma }\phi _2(\omega
_n^{\prime })\phi _1(\omega _n^{\prime },q)\right\} .  \label{np}
\end{eqnarray}

\begin{center}
---------------------------------------------------------------
\end{center}

\begin{multicols}{2}

In the above equations, $x=eV_s/\omega _s$, $\omega _n^{\prime }=\omega
+\omega _J+n\omega _s$, $\omega _n^{\prime \prime }=\omega +eV+n\omega _s$, $
C=-\sum_p{\Delta \varphi _p^2/}\left( {\varepsilon _p^2+\Delta ^2\varphi _p^2
}\right) $, $\rho _s$ is the density of states of the superconductor in the
normal state, and $\rho _n(\nu )$ is the density of states of the normal
lead. In Eq. (\ref{np}), 
\begin{eqnarray*}
\phi _1(\nu ,q) &=&\pi \sum_k\varphi _k^2\left( 1-n_k-n_{-k+q}\right) \delta
(\nu -\varepsilon _k-\varepsilon _{-k+q}), \\
\phi _2(\nu ,q) &=&-\sum_k\varphi _k^2{\frac{{1-n_k-n_{-k+q}}}{{\ \nu
-\varepsilon _k-\varepsilon _{-k+q}}},}
\end{eqnarray*}
and $n_k=1/(1+e^{\beta \varepsilon _k})$ is the Fermi-Dirac function .

\begin{figure}[htb]
\begin{center}
\psfig{file=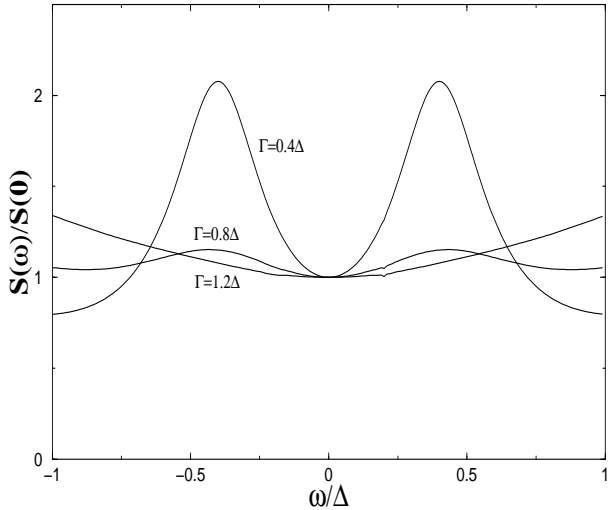,height=70mm,width=80mm,angle=-90}
\end{center}
\caption{The normalised current noise spectrum as a
function of the noise frequency for three different $\Gamma$. 
$\omega_J=0.4\Delta$.}
\label{Fig.2}
\end{figure}

Figure 2 shows $S(\omega )$ normalised by $S(0)$ as a function of $\omega $
for the case $\omega _J=0.4\Delta $ and $\omega _s=0$. The parameters used
are $\Delta _n=\Delta $, $T=0.2\Delta $, $V_0=2V_1=2\Delta $, and $\rho
_s=1/10\Delta $. The band width of the normal lead is assumed to be $
10\Delta $. From the figure, it is clear that if the fluctuation rate $
\Gamma $ is sufficiently small, the noise peak centred at $\omega =\pm
\omega _J$ is very sharp. The half width of the peak is determined by $
\Gamma $. When $\omega _J\gg \Gamma $, the thermally excited pair tunnelling
current will oscillate a few times before it loses the phase coherence. But
with increasing $\Gamma $, the noise peak at $\omega _J$ becomes broadened
and vanishes eventually.

\begin{figure}[htb]
\begin{center}
\psfig{file=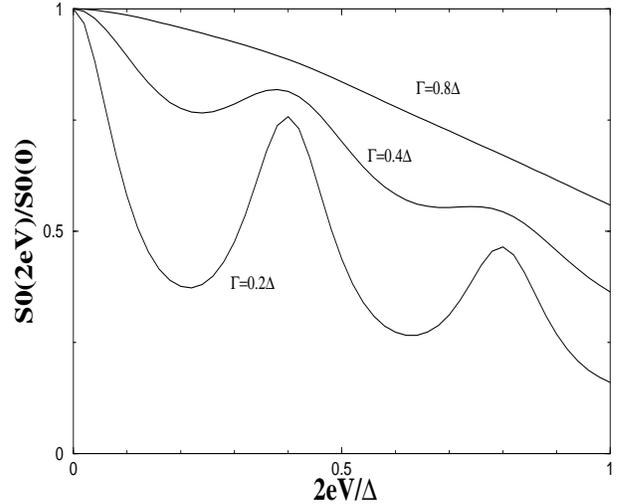,height=70mm,width=80mm,angle=-90}
\end{center}
\caption{The normalised current noise spectrum at zero noise 
frequency as a function of the intrinsic
Josephson frequency $\omega_J=2eV$ for three different $\Gamma$. 
$\omega_s = 0.4\Delta$. }
\label{Fig.3}
\end{figure}

In the presence of both DC and AC voltages, the pair tunnelling is enhanced
at the frequency $\omega =\omega _J+n\omega _s$ and the noise spectrum shows
a serial of local maxima at these frequencies. If the applied microwave
frequency and the Josephson frequency satisfy the condition $\omega
_J+n\omega _s=0$, a local maximum will appear at the zero frequency. Figure
3 shows the zero frquency 
normalised current noise spectrum as a function of $\omega _J$
for the case $\omega _s=0.4\Delta $ (the other parameters are the same as in
Figure 2). There are a series of peaks for $\Gamma =0.2\Delta $. When $
\Gamma $ is increased, these peaks becomes broader and weaker.

The above results indicate that the pair fluctuation in the normal 
lead has strong effects on the noise spectrum of the tunnelling current. 
The broadened peaks at the characteristic frequencies in the noise spectrum 
shown in Figures 2 and 3 are unique to the normal-superconductor junction 
with strong pair fluctuations. Experimentally, it is not easy
to measure the finite
frequency noise, but when the AC voltage is applied we only need
to measure the zero frequency noise which is quite easy to do.
Thus experimental measurements of the current 
noise allow us to judge how strong the pairing fluctuation is in the pseudogap 
phase. This is apparently important for a direct test of the superconducting 
precursor scenario.

In summary, we have shown that the noise spectrum of the 
tunnelling current between a underdoped cuprate in the normal state and 
an optimally doped cuprate in the superconducting state is a useful 
probe of the pairing fluctuation in the pseudogap phase. 
We have calculated the noise spectrum using the closed path Green's
functions and found that the current noise is strongly enhanced at 
a series of characteristic frequencies determined by the 
Josephson pair tunnelling processes. Notice that the system is 
nonequilibrium when the DC voltage is applied, it is proper to use
the closed time green's function technique.
Experimental observations of the noise peaks at these characteristic 
frequencies will help us to establish the correct picture of the 
pair fluctuation in the pseudogap phase of high-$T_c$
cuprates.

TX acknowledges the financial
support from the National Natural Science Foundation of China and from the 
special funds for Major State Basic Research Projects of China.

\end{multicols}

\end{document}